\documentclass[final,jgrga]{agutex}
\authorrunninghead{STOCKWELL AND COX}
\usepackage{graphicx}
\titlerunninghead{INFLUENCE OF THE SOUTHERN OSCILLATION}
\begin{document}

\title{Comment on "Influence of the Southern Oscillation on tropospheric temperature" by J. D. McLean, C. R. de Freitas, and R. M. Carter} 

\authors{David R.B. Stockwell, \altaffilmark{1} and Anthony Cox} 
\altaffiltext{1}{http://landshape.org, Email: davids99us@gmail.com}

\begin{abstract}

We demonstrate an alternative correlation between the El Ni\~no Southern Oscillation (ENSO) and global temperature variation to that shown by \cite{McLean:2009ly}. We show 50\% of the variation in RATPAC-A tropospheric temperature (and 54\% of HadCRUT3) is explained by a novel cumulative Southern Oscillation Index (cSOI) term in a simple linear regression.  We review evidence from physical and statistical research in support of the hypothesis that accumulation of the effects of ENSO can produce natural multi-decadal warming trends. Although it is not possible to reliably determine the relative contribution of anthropogenic forcing and SOI accumulation from multiple regression models due to collinearity, these analyses suggest that an accumulation ratio cSOI/SOI of $4.8\pm1.5\%$ and up to $9\pm2\%$ is sufficient for ENSO to play a large part in the global mean temperature trend since 1960.

\end{abstract}

\begin{article}

\section{Introduction}

\cite{McLean:2009ly} claim a high correlation between the Southern Oscillation Index (SOI) and tropospheric temperature on lagged, detrended data. These correlations suggest the involvement of the El Ni\~no Southern Oscillation (ENSO) in global temperature variations, but not warming trends due to the various detrending operations performed in the course of analysis.    \cite{G.-Foster:2009qd} question their approach, also citing research to date indicating that SOI explains between 15 and 30\% of interannual variability in surface and/or lower tropospheric temperature, but little of the global mean warming trend, generally attributed to an increase in anthropogenic greenhouse gas concentrations \citep{IPCC:2007ve}.  We bring attention to another suggestion by Bob Tisdale (http://bobtisdale.blogspot.com/) for explaining warming trends via a cumulative function of the SOI (cSOI), representing a residual effect of ENSO events on global temperature, with potential to account for a large part of the increase in global temperature since 1960.  

The increase in the frequency and intensity of ENSO since the mid-1970s compared with the previous 100 years may have increased global surface temperature during the last few decades, although the magnitude of the effect is thought to be small \citep{IPCC:2001ct}.  Recent results show a radiation imbalance related to the Pacific Decadal Oscillation (PDO) of $+0.15 W/m^2$ over the period \citep{Douglass:2009dq}.  Various lines of research provide physical explanations for globally persistent ENSO-effects \citep{Jacobs:1994sf, White:2003rz}.  In particular, the El Ni\~no events produce larger sea surface anomalies more than La Ni\~na \citep{Monahan} and increase solar absorption \citep{pavlakis}.  El Ni\~no effects may be due intensified by positive feedbacks such as relaxing of the trade winds further increasing heating  \citep{Sun:2009fp}, or strong positive feedback from low level clouds over the Pacific \citep{Clement:2009rr}.   \cite{Sun:2009fp} propose a dynamic whereby the asymmetric warming leads to a slow increase in the Pacific mean temperature during periods of persistent El Ni\~no, ultimately leading to a reversion back to more neutral La Ni\~na conditions.  Thus, it has been speculated that global warming on decadal, interdecadal and even centennial scales may be due to internal modes of variability \citep{White:2003rz}.  The potential for a natural source of warming during the 1960-1998 period is supported by records of decreasing cloud cover during the El Ni\~no dominated positive phase of the Pacific Decadal Oscillation \citep{Wielicki:2002vn, Palle:2005zr}.   

Purely statistical analysis suggests a (nearly) unit-root (or random walk) property in the global temperature series \citep{Stockwell:2006rf, Kout:2008tg}.  It is well known that a random walk can be produced simply from the cumulative sum of an independent random variable, and although the source has not been identified, physically realistic versions of random walks, such as mean-reverting series, long-term persistent (LTP), or self-similar series, are suggested by global temperature trends \citep{Cohn:2005ys}.  The apparent unit-root property of global temperature may be caused ocean integration of ENSO effects.

Some climate models show multi-decadal natural variation in temperature, although attribution of these trends to ENSO may be confounded, as some GCMs show an ENSO-like response to global warming (although others show an enhanced Northern Annular Mode) \citep{IPCC:2007ve}.   Models with insufficient variability at decadal time scales are unlikely to show ENSO accumulation.  For example, a multilayer ocean model would be needed to represent a persistent re-emergence mechanism as proposed by \cite{Alexander:1995db}, whereby anomalies carry over from the previous years winter water through entrainment into the mixed layer as it deepens the following winter \citep{Timlin:2002hl}.   Clouds are still acknowledged as the major source of uncertainty, and inadequacies in the tropical energy budget such as the dynamics and variability of top-of-atmosphere radiation \citep{Wielicki:2002vn} and lack of realism in the handling of the tropical precipitation relations and transfer of heat \citep{Jungclaus:2006gf} might also impede recognition.  Thus while models are broadly consistent in attributing recent temperature trends to anthropogenic forcing, they do not rule out the potential for internal variability to account for a large part of the increase in temperature from 1960 to 2008 \citep{Wigley:1995rm}.

\section{Analysis}

The cumulative SOI as a function of time $t$ is:

\[cSOI(t) = \sum_{t=1}^{t} SOI_t\]

We fit a series of multiple regression models to 12-month filtered (or smoothed) tropospheric global temperature series as per \cite{McLean:2009ly} (RATPAC-A available at http://www1.ncdc.noaa.gov from \cite{NOAA:2009ly}), and to an unfiltered global land and ocean surface dataset (HadCRUT3 -- available at http://www.cru.uea.ac.uk by \cite{Jones:1999ar}).  The independent terms in the models were composed of the Southern Oscillation Index (SOI -- available at http://www.bom.gov.au/climate/current/soihtm1.shtml), the sunspot number as a proxy for solar irradiance (SSN -- available at http://solarscience.msfc.nasa.gov) and from the \cite{GISS:2009} website, the stratospheric aerosols (V -- available at http://data.giss.nasa.gov) and anthropogenic forcing from the smoothed sum of all variables other than volcanism and solar forcing (AF -- available at http://data.giss.nasa.gov).  The SOI was not lagged.

\begin{figure}[p]
\includegraphics{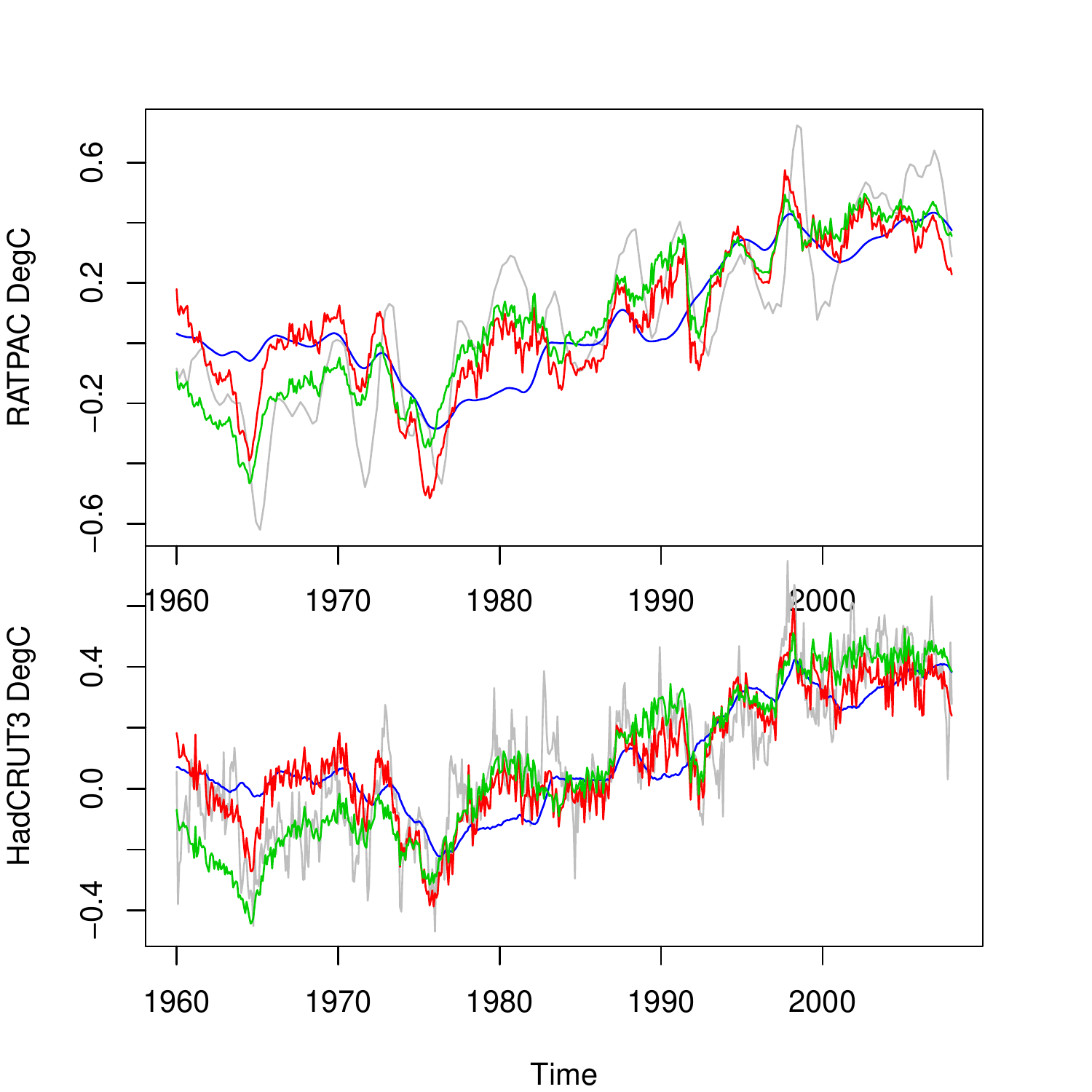}

\caption{The fit of models to temperature for RATPAC-A filtered (above)  and HadCRUT3 raw (below), including the global temperature (gray line), cumulative SOI only (blue line), the natural model (red line), and full model including anthropogenic forcing (green line).}
\label{fig1}
\end{figure}

Figure~\ref{fig1} shows various fitted models for RATPAC-A (upper, filtered as in \cite{McLean:2009ly}) and HadCRUT3 (lower).  The fit of cSOI (blue line) to temperature is apparent, and further enhanced with SOI, volcanics (V) and sunspot number (SSN) (red line).  Inclusion of the anthropogenic forcing term (AF) increases the overall fit slightly (green line).  Also worthy of note, the maximum and minimum of global temperatures (HadCRUT3) are coincident with maximum and minimum of cSOI over the same period.  

Table~\ref{tab1} lists the correlation results for the multiple regression models of global temperature.  The $R^2$ coefficients of determination (proportion of variation explained) for the in-sample (R2) and the split-segment cross-validation average (R2c, split on 1984) are listed for both RATPAC-A (RAT) (SOI and RATPAC 12-month filtered but not lagged), and HadCRUT3 (HAD) (not filtered or lagged) global surface temperature data sets. 

Referring to the results for RATPAC in Table~\ref{tab1}, model $GTTA\sim SOI$, though significant, explains a small proportion of the variation (7\%) from 1960 to 2008.  The addition of the cSOI term increases the variance explained to 50\% ($GTTA\sim cSOI$). The anthropogenic forcing model, $GTTA\sim AF$, explains 69\% of variation, increasing to 71\% when SOI terms are included.  The full model explains 78\% of the variation.  

The variation explained in the HadCRUT3 data (Table~\ref{tab1}) by cSOI is a larger 54\%, and the inclusion of SOI terms improve the anthropogenic forcing model from 66\% to 72\%.  Finally, the full model explains 77\% of the variation.  The cross-validation $R^2$ results are lower but consistent with the in-sample validation results, demonstrating the models are robust to out-of-sample split-segment validation.

While one would like to know the relative contribution of anthropogenic forcing and SOI accumulation, reliable determination with multiple regression models is not possible due to the collinearity of the variables.  We can, however, calculate some values suggested by these analyses and with greater confidence estimate an upper bound on ENSO accumulated in the absence of anthropogenic forcing.  The coefficients for two equations (all very significant) were:

\begin{eqnarray*}
&T_1 & =   0.29\pm0.01 \\
	&& - 0.00055\pm0.00001 cSOI - 0.0061\pm0.0006 SOI   \\
	&&	- 1.8\pm0.2 V - 0.0011\pm0.0001 SSN   \\
&T_2 & =   -0.15\pm0.03 \\
	&&   - 0.00025\pm0.00002 cSOI - 0.0051\pm0.0005 SOI \\
	&&	- 1.6\pm0.2 V - 0.0007\pm0.0001 SSN  + 0.19\pm0.01AF 
\end{eqnarray*}

The accumulating proportion of SOI is estimated from the ratio of the coefficients of cSOI and SOI, and from the maximum ranges of temperature and cSOI.    The ratio of coefficients cSOI/SOI for HadCRUT3 ($T_1$) indicates that up to $9\pm2\%$ (95\%CL from coefficient uncertainty) of the SOI is accumulated. The equivalent ratio of coefficients cSOI/SOI for HadCRUT3 ($T_2$) including anthropogenic forcing (AF) suggests a value of $4.9\pm1.5\%$ may represent the fraction of SOI accumulated, however this figure needs to be viewed with caution.  

The cSOI ranged from 948 in 1976 to -172 in 1998 and global temperature ranged 0.8817 degrees Celsius between the coincident maximum and minimum over the same period.  Therefore the contribution of cSOI to the increase is potentially 0.8817/1120 or 0.00079 degrees of global temperature per SOI unit.  When divided by the SOI coefficient ($T_1$) this gives a higher estimate of $13\pm3\%$, as it incorporates some 'overshoot' at the extreme highs and lows of temperature.  

\section{Discussion}

These analyses suggest that a contribution from ENSO-effects to global temperatures, when expressed as the cumulative sum of the SOI, can potentially account for 50\% of the variation in global mean temperature in the last 50 years -- a 'large part' of warming, as claimed by \cite{McLean:2009ly}.  

We determine the relative contribution of ENSO and anthropogenic forcing, with the qualification that linear regression studies are not reliable when variables are collinear, and other data are needed to determine whether both, either, or neither contribute in reality.  However, our estimated sensitivity of global mean temperature to anthropogenic forcing when cSOI is included is $0.19\pm0.02\% K/W/m^2$ (AF coefficient equation 2), which is smaller, though close to the no-feedback sensitivity value for a number of forcings (including $CO_2$) of $0.30 K/W/m^2$ \citep{douglass:2008}. As an estimated sensitivity of around $0.8 K/W/m^2$ is required for climate models to reproduce the observed warming trend since 1960, it follows that if the accumulated ENSO effect were to make up the difference in global mean trend since 1960, ENSO must be responsible for greater than 50\% of the observed warming.  Therefore, providing interactions between ENSO and greenhouse gas forcing are not significant, these results are clearly inconsistent with the claim that "[M]ost of the observed increase in global average temperatures since the mid-twentieth century is most likely due to the observed increase in anthropogenic greenhouse gas concentrations" \citep{IPCC:2007ve}.  The global temperature change from ENSO events that needs to be accumulated to achieve the observed trends is estimated from the ratio of the coefficients cSOI/SOI at $4.9\pm1.5\%$ and may be up to 9\%.  

These analyses show it would be premature to dismiss a large role for ENSO in the global mean temperature trend since 1960.  The cumulative SOI together with physical research and statistical modelling, provide support for the contention that multi-decadal natural variability from ENSO events may play a larger part in recent warming than is generally appreciated, although further studies of longer temporal duration and spatial extent are needed to better understand and quantify the mechanisms involved.

\bibliographystyle{agufull08}

 \end{article}
\newpage
\begin{table}[ht]
\begin{center}
\begin{tabular}{lcccc}
  \hline
 & RAT R2 & RAT R2c & HAD R2 & HAD R2c \\ 
  \hline
GTTA\~{}SOI & 0.07 & 0.06 & 0.08 & 0.10 \\ 
  GTTA\~{}cSOI & 0.50 & 0.22 & 0.54 & 0.23 \\ 
  GTTA\~{}AF & 0.69 & 0.44 & 0.66 & 0.37 \\ 
  GTTA\~{}cSOI+SOI+AF & 0.71 & 0.43 & 0.72 & 0.42 \\ 
  GTTA\~{}cSOI+SOI+V+SSN+AF & 0.78 & 0.52 & 0.77 & 0.51 \\ 
   \hline
\end{tabular}
\caption{In-sample (R2) and split-sample cross-validation (R2c) correlation coefficients of regressions of global mean temperature series RATPAC-A (RAT 12-month filtered) and HaDCRUT3 (HAD).}
\label{tab1}
\end{center}
\end{table}

\end{document}